\documentclass[aps,pra,10pt,twocolumn,groupedaddress,notitlepage,showpacs,floatfix,longbibliography]{revtex4-1}
\pdfoutput=1
\usepackage{graphicx,graphics,epsfig,subfigure,times,bm,bbm,amssymb,amsmath,amsfonts,amsthm,mathrsfs,MnSymbol}
\usepackage{gensymb}
\usepackage[matrix,frame,arrow]{xypic}
\usepackage[pdfstartview=FitH]{hyperref}
\usepackage{subfigure}
\hypersetup{
    colorlinks=true,       
    linkcolor=red,          
   citecolor=magenta,        
    filecolor=magenta,      
    urlcolor=cyan,           
    runcolor=cyan
}
\usepackage[pdftex]{color}
\usepackage{braket}
\usepackage{enumerate}
\usepackage[normalem]{ulem}
\usepackage[usenames,dvipsnames]{xcolor}
\usepackage{multirow}
\usepackage{mathtools}

\definecolor{orange}{rgb}{1,0.5,0}

\newcommand{\bes} {\begin{subequations}}
\newcommand{\ees} {\end{subequations}}
\newcommand{\bea} {\begin{eqnarray}}
\newcommand{\eea} {\end{eqnarray}}

\definecolor{gold}{rgb}{0.85,.66,0}

\newcommand{\beq}{\begin{equation}}

\newcommand{\eeq}{\end{equation}}

\newcommand{\ignore}[1]{}

\def\s{\sigma}

\def\>{\rangle}
\def\<{\langle}

\def\s0{I}

\newcommand{\ig}[1]{}

\begin{document}
\title{Catalyzed entanglement concentration of qubit pairs}
\author{Siddhartha Santra}
\affiliation{US Army Research Laboratory, Adelphi, Maryland 20783, USA}
\author{Vladimir S. Malinovsky}
\affiliation{US Army Research Laboratory, Adelphi, Maryland 20783, USA}
\begin{abstract}
We analytically obtain the maximum probability of converting a finite number of copies of an arbitrary two-qubit pure state to a single copy of a maximally entangled two-qubit pure state via entanglement assisted local operations and classical communications using a two-qubit catalyst state. We show that the optimal catalyst for this transformation is always more entangled than the initial state but any two-qubit state can act as a (non-optimal) catalyst. Interestingly, the entanglement of the optimal two-qubit catalyst state is shown to decrease with that of the initial state. Entanglement assisted strategies for obtaining multiple Bell states are discussed.
\end{abstract}
\maketitle


Entanglement concentration (EC) \cite{entconc1} is the process of obtaining maximally entangled pure states given some initial number of copies, $N$, of partially entangled pure states using local quantum operations and classical communications (LQCC) \cite{entmonotones1}. Concentrated entanglement is an important resource for applications \cite{santratelescope,wilde_2017, nielsen_chuang_2010} and EC protocols are of fundamental interest in quantum information theory \cite{wilde_2017, nielsen_chuang_2010}. Various LQCC EC protocols, which work for different numbers of initial states and with varying efficiencies, are known \cite{entconc1,lo-popescu,bose1}. Although LQCC is a natural operational paradigm, where observers Alice and Bob each possess and operate only on part of a quantum system while coordinating their actions through classical communications, more efficient EC protocols can be obtained using entanglement-assisted local quantum operations and classical communications (ELQCC) \cite{catalysis1, catalysis2}. In this process, an ancillary entangled pure state, called the catalyst state, shared by Alice and Bob is utilized as part of an overall LQCC process to enhance its efficiency and the catalyst state is recovered intact at the end. Here, we analytically obtain the maximum probability of success for an EC protocol transforming $N$-copies of a two-qubit pure state to a single copy of a maximally-entangled two-qubit pure state, or Bell state, when provided with entanglement assistance in the form of a two-qubit pure state catalyst. 

In the case of a large number of copies, $N\to \infty$, of a two-qubit pure state, $\ket{\alpha}=\sqrt{\alpha}\ket{00}+\sqrt{1-\alpha}\ket{11}$, a fundamental result \cite{entconc1} is that the number, $M$, of Bell states $\ket{\phi}=(\ket{00}+\ket{11})/\sqrt{2}$, obtainable using LQCC achieves the value, $M=S_{VN}(\alpha)N$, with $S_{VN}(\alpha)=-\alpha\log_2(\alpha)-(1-\alpha)\log_2(1-\alpha)$ - the Von Neumann entropy of the reduced initial state. The result is interpreted to mean that a fraction, $f=M/N$, of the initial states are deterministically transformed to Bell states. A single Bell state, in the limit of an asymptotic number of copies of $\ket{\alpha}$ with $\alpha\neq1$, can always be obtained with certainty. In the other limit, for $N=1$, a Bell state can be obtained only probabilistically via LQCC with the maximum probability being, $P=2(1-\alpha)<1$, \cite{lo-popescu,bose1} since without loss of generality $\alpha\geq0.5$. However, ELQCC does not increase the success probability of a transformation from a single copy of a two-qubit state to a Bell state. Therefore, in both these limits, i.e. $N\to1$ and $N\to\infty$, entanglement assistance does not help, that is, it cannot increase the number of Bell states, $M$, obtained asymptotically nor can the success probability, $P$, be increased for a single copy of $\ket{\alpha}$. However, in the intermediate regime of $N$, ELQCC can increase the expectation value of entanglement obtained in the form of maximally entangled states (of any dimension) in an EC procedure \cite{catalysis1}.

We show that for finite $N\geq 2$ entanglement assistance increases the success probability of the transformation, $\ket{\alpha}^{\otimes N}\to\ket{\phi}$. We analytically find that while all pure and entangled two-qubit states can act as catalysts for this tansformation, i.e. increase its success probability, the optimal catalyst must be more entangled than the initial state $\ket{\alpha}$. Remarkably, we find that the entanglement of the optimal catalyst decreases with that of the initial state. Further, we find that the use of an ELQCC procedure for EC is most beneficial for smaller number of copies, $N$, of the initial state. To close, we comment on ELQCC strategies to obtain multiple copies of Bell states. Obtaining catalysts for entanglement transformations is in general a difficult problem analytically while numerical searches do not provide much insight into the general properties of catalyst states. 

Entanglement assistance via the presence of a catalyst state, $\ket{C}$, can enable an otherwise impossible LQCC entanglement transformation \cite{catalysis1}, i.e.,
\begin{align}
\ket{\psi}&\underset{LQCC}{\not\to}\ket{\phi}\nonumber\\
\ket{\psi}\ket{C}&\underset{LQCC}{\to}  \ket{\phi}\ket{C}.
\end{align}
This result is based on Nielsen's theorem \cite{Nielsen1} which provides a criterion for allowed LQCC transformations from one pure quantum state to another. The criteria states that the transformation from an initial state $\ket{I}$ to a final state $\ket{F}$ is possible with certainty, i.e. $P(I\to F)=1$, iff the sets of the squares of the non-increasingly ordered Schmidt coefficients (OSC), $\bar{\lambda}^I=(\lambda^I_1\geq\lambda^I_2\geq...\geq\lambda^I_d)$ and $\bar{\lambda}^F=(\lambda^F_1\geq\lambda^F_2\geq...\geq\lambda^F_d)$ with respect to the bipartition that defines the local quantum systems, obey the majorization relation, 
\begin{align}
\bar{\lambda}^I\preceq \bar{\lambda}^F,
\label{majrel}
\end{align}
which is shorthand to denote that $\sum_{j=1}^k \lambda^I_j\leq \sum_{j=1}^k \lambda^F_j~\forall 1\leq k \leq d$. In case of incommensurate states, i.e. where the OSCs of the initial and final states do not obey Eq.~(\ref{majrel}), Vidal \cite{vidal1} showed that the transformation from $\ket{I}\to\ket{F}$ is possible only probabilistically with the maximum probability given by,
\begin{align}
P(I\to F)=\underset{1\leq l \leq d}{\text{min}}\frac{E_l(\ket{I})}{E_l(\ket{F})},
\label{LQCCprob}
\end{align}
where $E_l(\ket{I}):=1-\sum_{j=0}^{l-1}\lambda^I_j$ and $\lambda_0=0$. For a pair of incommensurate states, Ref. \cite{catalysis1} further showed that entanglement catalysis can increase the efficiency, $P_C(I\to F)>P(I\to F)$, of probabilistic transformations. It is this approach we take to obtain catalysts that can maximize the LQCC entanglement concentration success probability of a finite number of two-qubit pure states.

For the problem of entanglement concentration of multiple copies of 2-qubit pure states, $\ket{\alpha}=\sqrt{\alpha}\ket{00}+\sqrt{1-\alpha}\ket{11}$, we have the initial and final states of the form,
\begin{align}
\ket{\psi}&=\ket{\alpha}^{\otimes N}\nonumber\\
\ket{\phi}&=(\ket{00}+\ket{11})/\sqrt{2},
\label{stateform}
\end{align}
which will be provided entanglement assistance via the catalyst state $\ket{C}$. We will first consider a fixed number of copies, $N$, in the above. Nielsen's theorem applied to the state pair of the form in Eq. (\ref{stateform}) implies the following,\\

\noindent \textbf{Proposition}: If the states $\ket{\psi}\to\ket{\phi}$ are incommensurate, no catalyst can make the transformation deterministic.
\begin{proof} First, note that incompatibility arises iff $\lambda^I_1>\lambda^F_1$ since $\lambda^F_1+\lambda^F_2=1$ and $\sum _{j=1}^{k} \lambda^I_j\leq 1 \forall 1\leq k\leq d$. Thus, the OSCs of the product states $\ket{\psi}\ket{C}$ and $\ket{\phi}\ket{C}$ remain incompatible since their largest Schmidt coefficients follow, $\lambda^I_1 c_1 > \lambda^F_1 c_1$, whereas $\sum_{j=1}^k \lambda^I_j\leq \sum_{j=1}^k \lambda^F_j~\forall ~2\leq k \leq d$. Here, $c_1$ is the square of the largest Schmidt coefficient of $\ket{C}$.
\end{proof}
For a fixed $N\geq1$, we focus on transformations $\ket{\psi}^{\otimes N}\to\ket{\phi}$ that are not possible with certainty using LQCC. The OSCs of the two states form probability vectors of length $2^N$ and are given by,
\begin{align}
\bar{\lambda}^\psi&=\{\alpha^N\geq\alpha^{N-1}(1-\alpha)\geq\alpha^{N-2}(1-\alpha)^{2}\geq...\geq(1-\alpha)^N\}\nonumber\\
\bar{\lambda}^\phi&=\{0.5\geq0.5\geq0\geq...\geq0\}
\end{align}
where the Schmidt coefficients $\alpha^{N-p}(1-\alpha)^p$ of $\ket{\psi}$ have multiplicities of ${N \choose p}$ and $0.5\leq\alpha\leq1$.
The optimal success probability for such a transformation as given by Eq.~(\ref{LQCCprob}) is,
\begin{align}
P(\psi\to\phi)=\text{min}[1,2(1-\alpha^N)].
\end{align} 
For LQCC transformations that are probabilistic the minimum in the R.H.S. above is less than unity. Therefore, we have that $2(1-\alpha^N)<1\implies \alpha>(1/2)^{1/N}$.
For such states we would like to find a catalyst, $\ket{C}=\sqrt{c}\ket{00}+\sqrt{1-c}\ket{11}$, i.e. a pure state on a qubit pair that provides the largest boost to the success probability, $P_C(I\to F)$, of the transformation,
\begin{align}
\ket{I}=\ket{\psi}\ket{C}\underset{LQCC}{\to}\ket{F}=\ket{\phi}\ket{C}.
\end{align}
To obtain $P_C(I\to F)$, first we need to evaluate the terms in the R.H.S of Eq.~(\ref{LQCCprob}). This requires the OSCs of the initial and final states. The OSCs of the final state $\ket{F}$ are,
\begin{align}
\bar{\lambda}^F=\{0.5c,0.5c,0.5(1-c),0.5(1-c),0,...,0\},
\end{align}
with $0.5\leq c\leq 1$ where the zeros following the non-zero entries make the length of $\bar{\lambda}^F$ match the dimension of the initial state $\text{dim}(\ket{I})=2^N\times 2$. Now, we note that the minimization problem in Eq.~(\ref{LQCCprob}) is restricted to the first four values of $l$ since, $E_l(\ket{F})=0\forall l\geq 5$, and thus the ratios, $r_l(\alpha,c):=E_l(\ket{I})/E_l(\ket{F})=\infty$, for $l\geq 5$ do not contribute to the complexity of the minimization in our case. Therefore, only the first four monotones, $E_l$, of the initial and final states are required. These can be obtained if the first 3 entries of the OSCs of the initial and final states are known. For the final state (in the entire domain $c\in(0.5,1)$) we have that,
\begin{align}
E_1(\ket{F})&=1,\nonumber\\
E_2(\ket{F})&=1-c/2,\nonumber\\
E_3(\ket{F})&=1-c,\nonumber\\
E_4(\ket{F})&=(1-c)/2.
\label{Ef}
\end{align}
For the initial state $\ket{I}$, the OSCs can have the following two orderings (of relevance are the first three entries of each) based on the value of $c$ relative to $\alpha$,
\begin{small}
\begin{align}
\bar{\lambda}^{I_1}&=\{c\alpha^N>(1-c)\alpha^{N}> c\alpha^{N-1}(1-\alpha)>...>(1- c)(1-\alpha)^N\}
\end{align}
\end{small}
which holds for $0.5<c\leq \alpha$ whereas for $\alpha<c\leq 1$,
\begin{small}
\begin{align}
\bar{\lambda}^{I_2}&=\{c\alpha^N> c\alpha^{N-1}(1-\alpha)> (1-c)\alpha^{N}>...> (1-c)(1-\alpha)^N\}
\end{align}
\end{small}
where the first three entries of $\bar{\lambda}^{I_1}$ have multiplicities $1,1,N$, while the multiplicities for the ordered entries of $\bar{\lambda}^{I_2}$ is $1,N,1$ respectively. Thus, for the two parts of the domain for $c$ the monotones $E_l(\ket{I})$ of the initial state evaluate to, 
\begin{align}
E_1(\ket{I})&=1,~c\in(0.5,1)\nonumber\\
E_2(\ket{I})&=1-c\alpha^N,~c\in(0.5,1)\nonumber\\
E_3(\ket{I})&=\begin{cases}1-\alpha^N,~~0.5<c\leq\alpha\\1-c\alpha^{N-1},~\alpha<c<1\end{cases}\nonumber\\
E_4(\ket{I})&=\begin{cases}1-\alpha^N-c\alpha^{N-1}(1-\alpha),~0.5<c\leq\alpha\\1-c\alpha^{N-1}-c\alpha^{N-1}(1-\alpha),~\alpha<c<1\end{cases}
\label{Ei}
\end{align}

From Eqs.~(\ref{Ef}) and (\ref{Ei}) we have the four ratios of the entanglement monotones as functions of $\alpha, c$ and $N$,
\begin{align}
r_1(\alpha,c,N)&=1,~c\in(0.5,1)\nonumber\\
r_2(\alpha,c,N)&=\frac{1-c\alpha^N}{1-c/2},~c\in(0.5,1)\nonumber\\
r_3(\alpha,c,N)&=\begin{cases}\frac{1-\alpha^N}{1-c},~~0.5<c\leq\alpha\\\frac{1-c\alpha^{N-1}}{1-c},~\alpha<c<1\end{cases}\nonumber\\
r_4(\alpha,c,N)&=\begin{cases}\frac{2(1-\alpha^N-c\alpha^{N-1}(1-\alpha))}{1-c},~0.5<c\leq\alpha\\\frac{2(1-c\alpha^{N-1}-c\alpha^{N-1}(1-\alpha))}{1-c},~\alpha<c<1\end{cases}
\label{R}
\end{align}

\emph{Evaluation of the minimum among the ratios of entanglement monotones:}
First, note that for $N=1$ the minimum of the ratios in the above set of equations is given by, $r_4(\alpha,c,N)=2(1-\alpha)$, which is equal to the LQCC probability without a catalyst for all values of $0.5<c<1$. Thus, a catalyst cannot help increase the success probability of a LQCC transformation of a single copy of $\ket{\alpha}$ to $\ket{\phi}$. This is consistent with the fact that catalysis is impossible when the initial and final states are both two-qubit states \cite{catalysis1}.

For $N\geq 2$, the minimum of the ratios $r_l(\alpha,c,N)$ for $l=2,3,4$ determine the probability of a successful catalyzed conversion from $\ket{\psi}^{\otimes N}\to \ket{\phi}$ (since $r_1(\alpha,c,N)=1$). For this we use the derivatives and continuity properties of $r_2,r_3,r_4$ to determine the minimum among the three. It turns out that for all values of, $\alpha>(1/2)^{(1/N)}$, the function $r_2(\alpha,c,N)$ decreases with $c$ with its maximum value $r_2^{\text{max}}=(4/3)(1-\alpha^N/2)$ as $c$ approaches $0.5$. On the other hand, the function $r_3(\alpha,c,N)$ increases with $c$ in both parts of its domain. It is continuous across the domain boundary $c=\alpha$ and has a minimum value of $r_3^{\text{min}}=2(1-\alpha^N)$ as $c$ approaches $0.5$. The minimum value of $r_2(\alpha,c,N)$ is given by $r_2^{\text{min}}=2(1-\alpha^N)$ as $c$ approaches $1$ whereas the value of $r_3(\alpha,c,N)$ diverges as $c\to1$. Therefore, for fixed $\alpha,N$ the curves for $r_2(\alpha,c,N)$ and $r_3(\alpha,c,N)$ as a function of $c$ intersect in the domain $c\in(0.5,1)$. Further, note that $r_2^{\text{max}}\geq r_3^{\text{min}}$ for $\alpha\geq(1/2)^{(1/N)}$. Finally, the minimum of the ratios is never given by the value of the function $r_4(\alpha,c,N)$ in any part of the domain $c\in(0.5,1)$ as shown in the following.

For $c\leq\alpha$, one can show that $r_4(\alpha,c,N)\geq r_3(\alpha,c,N)$ for all $N\geq2$, so that $r_4(\alpha,c,N)$ is not the least of the ratios as follows,
\begin{align}
r_4(\alpha,c,N)&=\frac{2(1-\alpha^N-c\alpha^{N-1}(1-\alpha))}{1-c}\nonumber\\
&=\frac{1-\alpha^N}{1-c}+\frac{1-\alpha^N-2c\alpha^{N-1}(1-\alpha)}{1-c}\nonumber\\
&=r_3(\alpha,c,N)+\frac{p(\alpha,c)}{1-c}
\end{align}
Now we note that the function, $p(\alpha,c,N)=1-\alpha^N-2c\alpha^{N-1}(1-\alpha)$, is a decreasing function of $c$ since $\alpha,(1-\alpha)\geq0$. So w.r.t. $c$ the function takes its minimum value at $c=\alpha$ given by, $p_{\text{min},c}(\alpha)=1+\alpha^N(2\alpha-3)$. This minimum value decreases with $\alpha$ since the sign of the derivative $dp_{\text{min},c}(\alpha)/d\alpha<0$ for $\alpha<(3/2)\frac{N}{N+1}$ which always holds for $N\geq 2$. The minimum value with respect to both arguments is at $c=\alpha$ and $\alpha=1$ and is given by $p_{\text{min},c,\alpha}=0$. 

For $\alpha<c<1$, one can show that $r_4(\alpha,c,N)\geq r_3(\alpha,c,N)$ for $N\geq 3$ whereas $r_4(\alpha,c,N)\geq r_2(\alpha,c,N)$ for $N=2$, so that also in this region $r_4(\alpha,c,N)$ is not the least of the ratios as follows. From Eq. (\ref{R}) we have,
\begin{align}
r_4(\alpha,c,N)&=\frac{1-c\alpha^{N-1}}{1-c}+\frac{1-c\alpha^{N-1}-2c\alpha^{N-1}(1-\alpha)}{1-c}\nonumber\\
&=r_3(\alpha,c,N)+\frac{q(\alpha,c)}{1-c},
\end{align}
where the function, $q(\alpha,c,N)=1-c\alpha^{N-1}-2c\alpha^{N-1}(1-\alpha)=1-c\alpha^{N-1}(3-2\alpha)$, is a decreasing function of $c$. Therefore, the minimum of $q(\alpha,c,N)$ w.r.t. $c$ is at $c=1$ and is given by $q_{\text{min},c}(\alpha)=1+\alpha^{N-1}(2\alpha-3)$. This minimum value decreases with $\alpha$ if the derivative $dq_{\text{min},c}(\alpha)/dq<0$ which requires $\alpha\leq (3/2)(N-1)/N$ that always holds for $N\geq3$. The minimum of $q_{\text{min},c}(\alpha)$ is therefore at $\alpha=1$ given by $q_{\text{min},c,\alpha}=0$ for $N\geq3$. For $N=2$, we have that, $(3/2)(N-1)/N=3/4$, so $q_{\text{min},c}(\alpha)<0$ for $(3/4)<\alpha<1$. However, for this range of $\alpha$ and $N=2$, we can show $r_4(\alpha,c,N=2)\geq r_2(\alpha,c,N=2)$ by evaluating their difference,
\begin{align}
&r_4(\alpha,c,N=2)-r_2(\alpha,c,N=2)\nonumber\\
&~~~~~~~~~~~~=\frac{2[1+(\frac{2c^2-4c}{\alpha}+(3c-2c^2))\alpha^2]}{(1-c)(2-c)}\nonumber\\
&~~~~~~~~~~~~=\frac{s(\alpha,c)}{(1-c)(2-c)},
\end{align}
where, $s(\alpha,c)=2[1+(\frac{2c^2-4c}{\alpha}+(3c-2c^2))\alpha^2]$. Note that the term $(2c^2-4c)$ decreases with increasing $c \forall c<1$ while the term $(3c-2c^2)$ decreases with increasing $c$ for $(3/4)<c<1$. Therefore, the minimum value of $s(\alpha,c)$ in this range is at $c=1$ given by $s_{\text{min},c}(\alpha)=2(1-\alpha)^2$ which is always greater than or equal to zero.
~~~~~~~~~~~~~~$\square$

\begin{figure}
\centering
\includegraphics[width=\columnwidth]{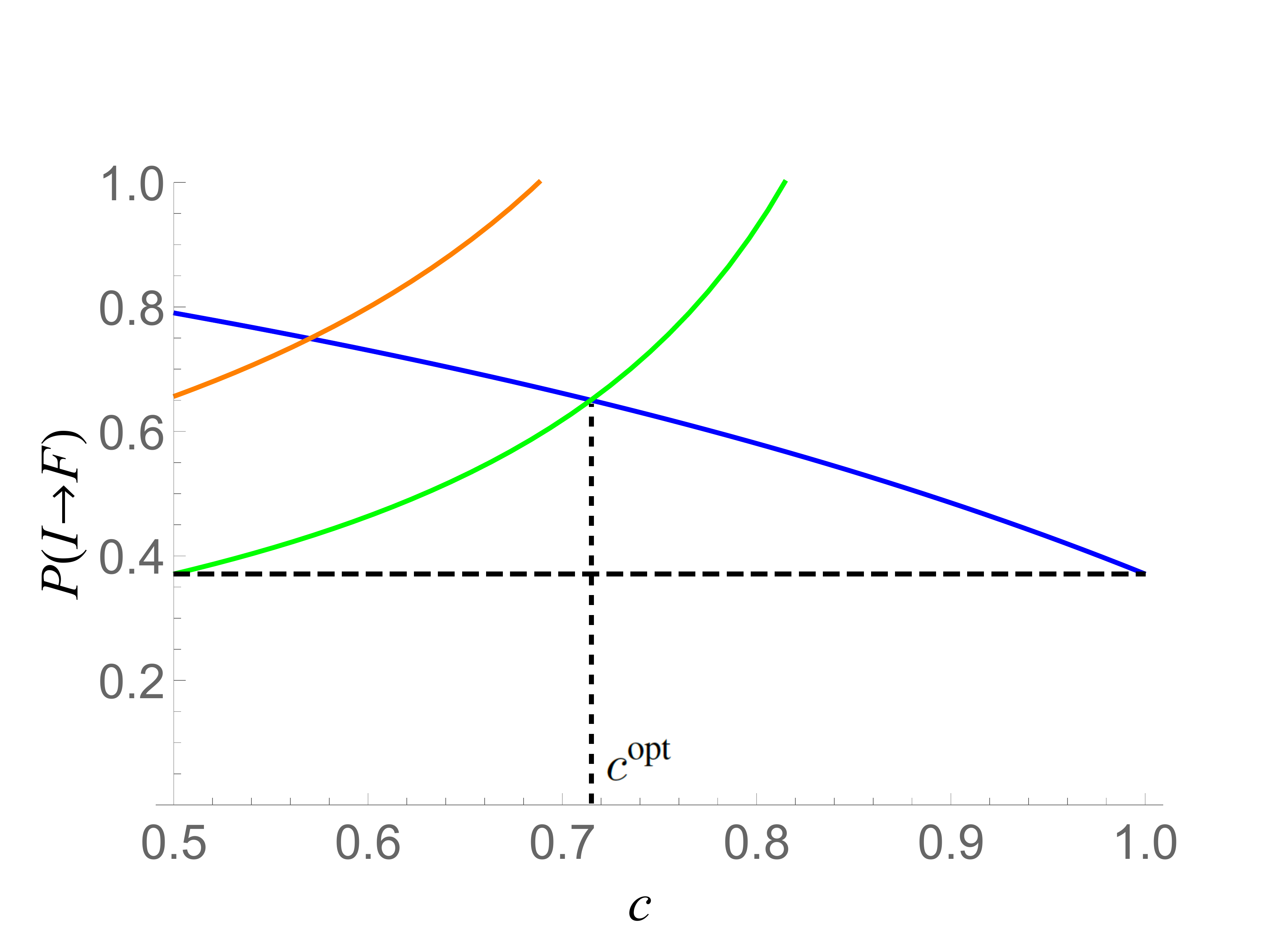}
\caption{Ratio of entanglement monotones as a function of the catalyst-state Schmidt coefficient, $c$, with fixed $\alpha=0.85$ and $N=2$. Shown in Blue is $r_2(0.85,c,2)$ which monotonically decreases while $r_3(0.85,c,2)$, in Green, monotonically increases with $c$. $r_4(0.85,c,2)$ shown in Red is never the minimum of the three monotones. The value of $c$ at the intersection point of the Blue and Green curves gives the optimal catalyst (vertical dashed line). The horizontal dashed line shows the probability for the LQCC transformation $\ket{\alpha=0.85}^{\otimes 2}\to\ket{\phi}$.}
\label{fig:optimalc1}
\end{figure}

These facts together imply that the maximum probability of a LQCC conversion, $\ket{I}\to\ket{F}$, is obtained where the curves for $r_2(\alpha,c,N)$ and $r_3(\alpha,c,N)$ w.r.t. $c$ intersect for a fixed $\alpha$ and $N$, see figure~(\ref{fig:optimalc1}). The intersection point,  $c^{\text{opt}}(\alpha,N)$, is obtained from the solution of one of the quadratic equations, $r_2(\alpha,c,N)=r_3^{c\leq\alpha}(\alpha,c,N)$, or, $r_2(\alpha,c,N)=r_3^{c>\alpha}(\alpha,c,N)$, as given by Eq.~(\ref{R}). We find that the latter has solutions, $c=0$ or $c>1$, which are unacceptable for a physically meaningful catalyst state, whereas the former equation provides an acceptable solution,

\begin{align}
c^{\text{opt}}(\alpha,N)=\frac{1+3\alpha^N-\{(1+3\alpha^N)^2-16\alpha^{2N}\}^{1/2}}{4\alpha^N}.
\label{catalystsolution} 
\end{align}

The Schmidt coefficient, $c^{\text{opt}}(\alpha,N)$, identifies a two-qubit catalyst pure state, $\ket{C^{\text{opt}}(\alpha,N)}=\sqrt{c^{\text{opt}}(\alpha,N)}\ket{00}+\sqrt{1-c^{\text{opt}}(\alpha,N)}\ket{11}$, that provides the maximum success probability in an ELQCC procedure to obtain a maximally entangled two-qubit state from $N$-copies of partially entangled pure states. This probability is given by the value of $r_2(\alpha,c^{\text{opt}},N)$ or $r_3(\alpha,c^{\text{opt}},N)$,
\begin{align}
P^{\text{max}}_{C}(I\to F)= \frac{1-\alpha^N}{1-c^{\text{opt}}(\alpha,N)}
\label{probvalue}
\end{align}

Further, since $c^{\text{opt}}(\alpha,N)<\alpha$ the optimal catalyst state is always more entangled than $\ket{\alpha}$. However, even those states, $\ket{C}=\sqrt{c}\ket{00}+\sqrt{1-c}\ket{11}$, with  $c\neq c^{\text{opt}}(\alpha,N)$ can act as (non-optimal) catalysts. This is because for such states $\ket{C}$ in the region $c<c^{\text{opt}}(\alpha,N)$ the minimum of the ratios, $r_3(\alpha,c,N)$, is still greater than the LQCC transformation probability of $2(1-\alpha^N)$ as can be seen by evaluating $r_3(\alpha,c,N)$ for $c<\alpha$, see the Green curve in figure~(\ref{fig:optimalc1}). Whereas for those states in the region $c>c^{\text{opt}}(\alpha,N)$ the minimum of the ratios, $r_2(\alpha,c,N)$, is again greater than the LQCC transformation probability of $2(1-\alpha^N)$, see the Blue curve in the same figure.

We remark that the transformation $\ket{I}\to\ket{F}$ can be achieved via LOCC operations jointly on the $N$-copies of the initial state and one-copy of the catalyst state in a two step procedure \cite{Nielsen1,vidal1,lo-popescu} we briefly outline. In the first step a temporary state $\ket{\Gamma}$ that majorises the initial state is obtained with certainty, i.e., $\ket{I}\prec\ket{\Gamma}$, via a sequence of LOCC operations on corresponding two-dimensional subspaces of Alice's and Bob's systems (of Hilbert space dimension $2^{N+1}$ each). That is, a single LOCC operation involves two-levels $\ket{i}_A,\ket{j}_A$ on Alice's systems and the corresponding two levels $\ket{i}_B,\ket{j}_B$ of Bob's systems with $i,j\in[1,2^{N+1}]$. Note that the operations on states, $\{\ket{i}_{A,B}\}_i$, involve the collective manipulation of $N$-qubits of the shared initial state and $1$-qubit of the shared catalyst state. The number of such $(\alpha,c)$-dependent two-level operations is upper bounded by $(2^{N+1}-1)$. In the second step, Bob performs a two-outcome generalized measurement on his portion of the shared state $\ket{\Gamma}$. For one of the outcomes, which occurs with probability given by Eq.~(\ref{probvalue}), the post-measurement state obtained is $\ket{F}$ therefore in this case the catalyst state is recovered along with a Bell state whereas the other outcome signals the failure of the catalytic process and the post-measurement state may be discarded.

\begin{figure}
\centering
\includegraphics[width=\columnwidth]{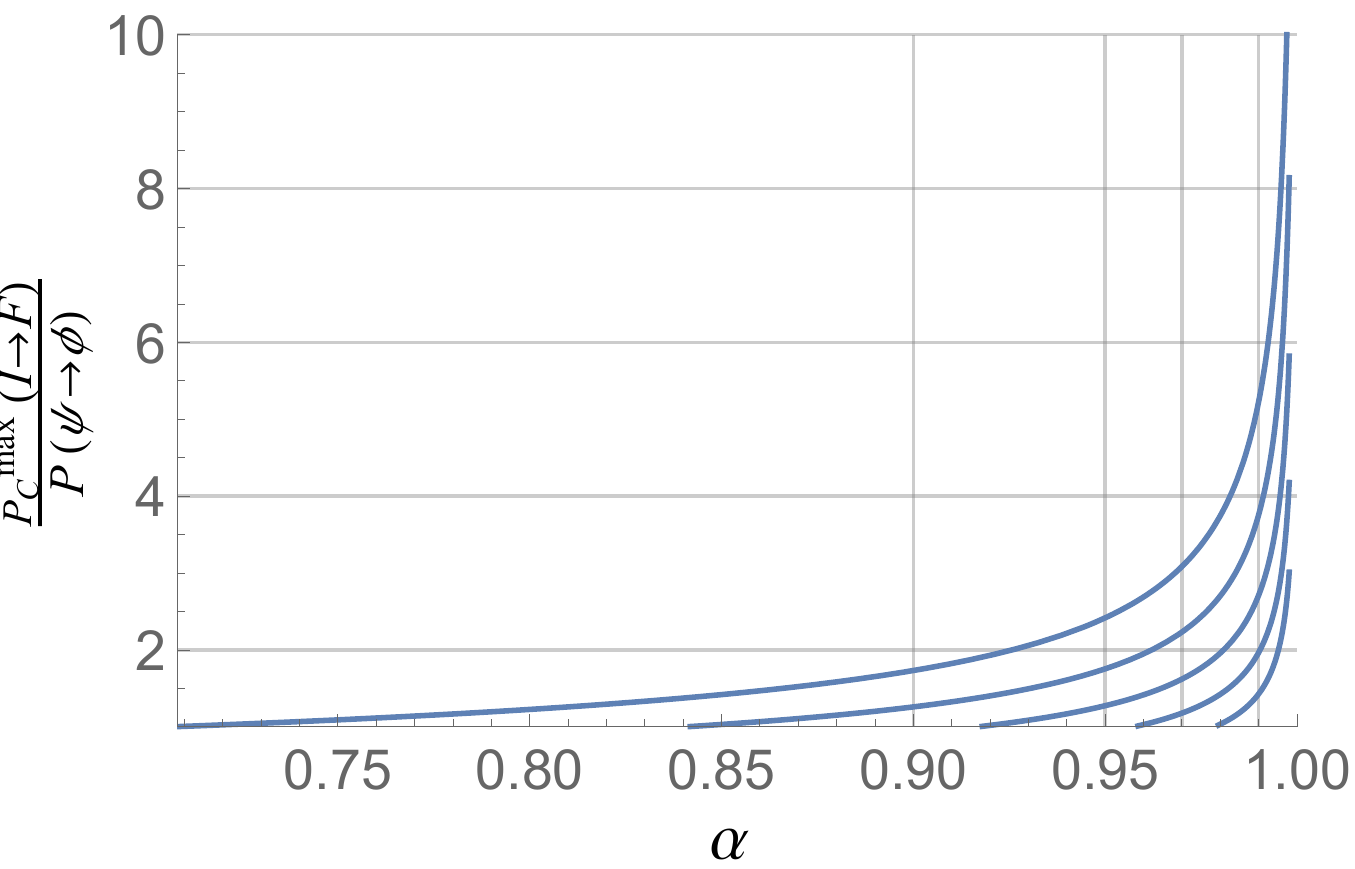}
\caption{Ratio of success probability for the transformation, $\ket{\psi}^{\otimes N} \to \ket{\phi}$, using optimal catalysis to that without catalysis. The curves from left to right are for different number of copies $N=2,4,8,16,32$.}
\label{fig:ratio1}
\end{figure}

Now, we note from Eqs.~(\ref{R}) and (\ref{catalystsolution}) the following properties,
\begin{enumerate}
\item An optimal two-qubit catalyst state always exists for $N\geq2$-copies of every state $\ket{\alpha}$ with $\alpha\in((1/2)^{1/N},1)$. 
\item The optimal catalyst state is always more entangled than $\ket{\alpha}$ since $c^{\text{opt}}(\alpha,N)< \alpha$. 
\item {\emph Any} pure and entangled two-qubit state can act as a catalyst, that is, it provides a positive boost to the success probability of the $\ket{\psi}^{\otimes N}\to\ket{\phi}, N\geq2$ transformation in an entanglement assisted procedure.
\item \emph{Optimal} self-catalysis is not possible, that is, $c^{\text{opt}}(\alpha,N)\neq\alpha$ for any $N$ and $\alpha<1$. However, an additional copy of the state $\ket{\alpha}$ can act as a non-optimal catalyst.
\item The optimal catalyst state $\ket{C^{\text{opt}}(\alpha,N)}$ becomes less entangled as the state $\ket{\alpha}$ becomes less entangled ($\alpha\to1$) since the derivative, $dc^{\text{opt}}(\alpha,N)/d\alpha>0$, in the region $\alpha\in(0.5,1)\forall N\geq2$.
\item Catalysis with the optimal state is more beneficial if the initial state is less entangled, that is, the ratio of LQCC success probability with optimal catalysis to that without, $\frac{P_C^{\text{max}}(I\to F)}{P(\psi\to\phi)}$ increases as $\alpha\to 1$, see figure~(\ref{fig:ratio1}).
\item Catalysis with the optimal two-qubit catalyst state is more effective for a smaller number of copies, $N$, of the initial state, see figure (\ref{fig:ratio1}).
\end{enumerate}

As a consequence of remark 3, we note that for a set of two-qubit pure states, $\mathcal{S}=\{\ket{\alpha_i}\}_i$ (none of which is a maximally entangled state), any two-qubit pure state can act as a common catalyst for all transformations, 
\begin{align}
\ket{\alpha_i}^{\otimes N_i}\to\ket{\phi},~N_i\geq2.
\end{align}

For obtaining multiple copies of Bell states higher dimensional catalysts are more efficient \footnote{Numerically one finds higher dimensional catalysts more efficient even to obtain a single Bell state. For example $\ket{C}=\sqrt{.5}\ket{00}+\sqrt{.35}\ket{11}+\sqrt{.15}\ket{22}$ is more efficient for the conversion $(\sqrt{.8}\ket{00}+\sqrt{.2}\ket{11})^{\otimes2}\to\ket{\phi}$}. For example, the initial state $\ket{\alpha}^{\otimes N}$ (with even $N$) can be transformed to $\ket{\phi}^{\otimes m}$ with a catalyst of the form $\ket{C^{\text{opt}}(\alpha,2)}^{\otimes N/2}$ in a pairwise ELQCC procedure where the number of obtained Bell states, $m=0,1,2,..,n=N/2$, is binomially distributed. The probability of obtaining $m$ Bell states is given by $p_m=\binom{N/2}{m}p^m(1-p)^{N/2-m}$ with $p=P^{\text{max}}_{C}(I\to F)$ as in Eq.~(\ref{probvalue}), where $\ket{I}=\ket{\alpha}^{\otimes 2}\ket{C^{\text{opt}}(\alpha,2)}$ and $\ket{F}=\ket{\phi}\ket{C^{\text{opt}}(\alpha,2)}$. The expected entanglement, $\braket{E}=\sum_m p_m*m=(N/2)P^{\text{max}}_{C}(I\to F)$, in this entanglement concentration procedure, that we will call strategy-1, is linear in the number of copies $N$ of the initial state $\ket{\alpha}$.

To obtain a target number, $m_*$, of Bell states, however, a different method, strategy-2, may be more beneficial. In such a strategy, the initial $N$-copies of $\ket{\alpha}$ may be grouped into $m_*$ sets each of cardinality $N_j$ such that, $\sum_{j=1}^{j=m_*}N_j=N$. The probability of obtaining $m_*$ Bell states will then be the maximum of the product of probabilities maximized over the size of the sets, $p_{m_*}=\text{Max}_{{\{N_j\}}_j}\prod_{j=1}^{j=m_*}P_j$, where $P_j$ is the probability of the transformation, $\ket{\alpha}^{\otimes N_j}\to\ket{\phi}$. For sets with $N_j\geq 2$ one can use an ELQCC transformation procedure, so that for such sets $P_j=P^{\text{max}}_{C}(I\to F)$ with $\ket{I}=\ket{\alpha}^{\otimes N_j}\ket{C^{\text{opt}}(\alpha,N_j)}$ and $\ket{F}=\ket{\phi}\ket{C^{\text{opt}}(\alpha,N_j)}$. The different cardinalities, $N_j$, of the sets allows one to maximize the catalysis success probability using the appropriate catalyst $\ket{C^{\text{opt}}(\alpha,N_j)}$ for each set. 

The choice of the advantageous strategy depends on the number of copies available $N$, the value of $\alpha$ and the number of copies of the Bell state $m_*$ desired as the output of the catalyzed entanglement concentration procedure. To compare, strategies-1 and 2 as described above, consider as an example the case when $N=6$ and $\alpha=0.99$. If $m
_*=2$ copies of Bell states are desired as output then strategy-1 yields a probability of $0.034$ whereas strategy-2 utilizing 2 sets of 3-copies of $\ket{\alpha}$ each, yields a probability of $0.065$. On the other hand if only a single copy of a Bell state is the desired output, i.e. $m_*=1$, then strategy-1 yields a probability of $0.391$ whereas strategy-2 utilizing 1 set of 6-copies of $\ket{\alpha}$ each yields a probability of $0.362$.

It will be interesting to apply the results of catalytic entanglement concentration to increase the efficiency of entanglement distribution protocols in quantum repeaters \cite{munro_repeater}. The latter distribute entanglement over long distances by purifying and connecting entanglement generated over smaller length segments. While the entanglement generated over the segments is typically in the form of mixed states, some models of channel noise, e.g. \cite{kwiat_filtration}, can lead to non-maximally entangled shared pure states between the repeater stations. In such cases, if ELQCC is utilized to extract states with high fidelity to a Bell state in each repeater segment more efficiently than LOCC based repeater protocols then the overall distribution rate of the repeater can benefit significantly. This would require adaptive operations at the repeater nodes since the transformation $\ket{I}\to\ket{F}$ is achieved via $\alpha$-dependent local unitaries and measurements by Alice and Bob. Copies of the initial states may be generated and stored on matter qubits that have an efficient light-matter interface while storing the catalyst state in long-lived quantum memories \cite{Simon2010} at the repeater nodes during the ELQCC process. This may allow the reuse of the catalyst state multiple times as allowed by the transformation success probability. Quantum repeater architectures based on the combination of qubits with excellent communication properties and those with long lifetimes, e.g. \cite{Santra_20192}, can thus be good candidates to exploit catalytic entanglement concentration.

In summary, we analytically obtained a two-qubit catalyst pure state that maximizes the success probability of an entanglement assisted LQCC procedure to convert a given number of copies of a partially entangled pure state to a single copy of a maximally entangled two-qubit state. The supplied entanglement assistance is minimal since the catalyst is an entangled state of Schmidt rank equal to 2. Although, a higher rank catalyst cannot make the transformation deterministic, the maximum transformation success probability with a catalyst of any rank is an open question. In contrast with numerical searches for catalyst states, the analytical derivation of the optimal catalyst state reveals multiple properties of the catalytic process and raises interesting questions about possible applications.

{\it Acknowledgements:-} We thank one anonymous referee for many useful comments and suggestions.

\bibliographystyle{apsrev4-1}
\bibliography{refs-greedy1}

\end{document}